\documentclass[manuscript]{acmart}

\usepackage{listings}
\usepackage{graphicx}
\usepackage{hyperref}
\usepackage{pgfplots}
\pgfplotsset{width=0.75\textwidth}

\AtBeginDocument{%
  }

\begin{document}

\title[Bounded Model Checking of RISC-V Machine Code with CFLOBDDs]{Bounded Model Checking of RISC-V Machine Code with Context-Free-Language Ordered Binary Decision Diagrams}


\author{Anna Bolotina}
\affiliation{%
  \institution{University of Salzburg}
  \city{Salzburg}
  \country{Austria}}
\email{anna.bolotina@plus.ac.at}

\author{Christoph M.~Kirsch}
\affiliation{%
  \institution{University of Salzburg}
  \city{Salzburg}
  \country{Austria}}
\affiliation{%
  \institution{Czech Technical University}
  \city{Prague}
  \country{Czechia}}
\email{ck@cs.uni-salzburg.at}

\author{Stefanie Muroya Lei}
\affiliation{%
  \institution{IST Austria}
  \city{Maria Gugging}
  \country{Austria}}
\email{stefanie.muroya@ist.ac.at}

\author{Matthias Pleschinger}
\affiliation{%
  \institution{University of Salzburg}
  \city{Salzburg}
  \country{Austria}}
\email{matthias.pleschinger@stud.plus.ac.at}


\begin{abstract}
Symbolic execution is a powerful technique for analyzing the behavior of software yet scalability remains a challenge due to state explosion in control and data flow. Existing tools typically aim at managing control flow internally, often at the expense of completeness, while offloading reasoning over data flow to SMT solvers. Moreover, reasoning typically happens on source code or intermediate representation level to leverage structural information, making machine code generation part of the trust base. We are interested in changing the equation in two non-trivial ways: pushing reasoning down to machine code level, and then offloading reasoning entirely into SMT solvers and other, possibly more efficient solver technology. In more abstract terms, we are asking if bit-precise reasoning technology can be made scalable on software, and not just hardware. For this purpose, we developed two tools called rotor and bitme for model generation and bounded model checking, respectively. We chose RISC-V restricted to integer arithmetic as modeling target for rotor since RISC-V integer semantics is essentially equivalent to established SMT semantics over bitvectors and arrays of bitvectors. While state-of-the-art SMT solvers struggle in our experiments, we have evidence that there is potential for improvement. To show the potential, we have slightly generalized and then implemented in bitme two types of binary decision diagrams (BDDs): algebraic decision diagrams (ADDs) and context-free-language ordered binary decision diagrams (CFLOBDDs). Bitme uses BDDs to propagate program input through models, essentially generalizing constant propagation to domain propagation. SMT solvers only get involved when model input cannot be propagated, significanly speeding up SMT solving. In other words, BDDs enable bitme to apply integer arithmetic natively rather than relying on SMT solving. We then study the impact on state explosion of CFLOBDDs, which are potentially more scalable than ADDs.
\end{abstract}

\maketitle

\section{Introduction}

We are interested in the problem of bit-precise reasoning over RISC-V machines with integer arithmetic using bitvector and array-of-bitvector logic. We have developed two tools: rotor, for generating bit-precise models of RISC-V~\cite{waterman2017risc} machines, and bitme, for reasoning over rotor-generated models using bounded model checking. Bitme utilizes Z3~\cite{de2008z3} and Bitwuzla~\cite{niemetz2023bitwuzla}, two state-of-the-art SMT solvers~\cite{brummayer2009boolector, niemetz2014boolector, niemetz2018btor2, de2008z3, niemetz2023bitwuzla}, as well as algebraic decision diagrams (ADDs)~\cite{frohm1993algebraic} and context-free-language ordered binary decision diagrams (CFLOBDDs)~\cite{sistla2024cflobdds, zhi2025polynomial}, both specifically tailored to bitvector logic as it appears in RISC-V machines.

Rotor generates models of 64-bit and 32-bit RISC-V machines without any loaded code, for code synthesis, as well as with code loaded into models, for code analysis through symbolic execution~\cite{cadar2008klee} on all execution paths simultaneously. Rotor can handle code generated by gcc as well as a custom-made non-optimizing compiler for source code limited to integer arithmetic. Model generation and size is linear in code size, hence constant without code. Intuitively, models allow checking safety properties~\cite{alpern1985defining} such as division-by-zero, non-zero exit code, and segmentation faults that are satisfiable if and only if there is machine input (program input and synthesized code) for which the modeled machine violates a safety property after executing a finite number of machine instructions. Machine input can be derived from satisfying assignments of model variables. In short, rotor models are sound and complete at bit level.

Bitme explores rotor-generated models breadth-first on all execution paths, that is, data flow and control flow simultaneously, using Z3 and bitwuzla as well as ADDs and CFLOBDDs. The BDDs~\cite{bryant1986graph} enable bitme to generalize constant propagation to domain propagation~\cite{schulte2005bounds} of all or some machine input through models while BDDs keep track of the relationship between machine input and current machine state. During model exploration bitme falls back to Z3 and bitwuzla if some machine input cannot be propagated. Bitme is sound and complete at bit level up to a given bound on number of executed instructions.

We use rotor and bitme as platform for studying and improving the performance of bit-precise reasoning technology applied to machine code as is, in contrast to hardware in particular, without using any higher-level program analysis techniques. Rotor can be seen as benchmark generator for applying bit-precise reasoning tools to source code that compiles to RISC-V while bitme enables comparing SMT solver performance on rotor-generated models with other bit-precise reasoning techniques such as BDDs.

The key challenge is to handle state explosion in control and data flow simultaneously, in particular without giving up on completeness. With CFLOBDDs we are particularly interested in machine code whose state space fluctuates in size during execution. The question is then whether CFLOBDDs are able to reflect that fluctuation by reusing CFLOBDD instances seen before to avoid exponential explosion in time and space.

\section{Contributions}

Rotor and bitme provide a platform for studying the question of whether bit-precise reasoning technology can be made to scale on models of software given in the form of machine code. The paper makes two key contributions:

\begin{enumerate}
\item rotor: efficient generation of RISC-V machine models that enable SMT solvers and other bit-precise reasoning technology to reason about unmodified RISC-V machine code, and
\item bitme: integration of potentially scalable BDD technology (CFLOBDDs) into a bit-precise reasoning engine for rotor-generated RISC-V machine models.
\end{enumerate}

Rotor and bitme are inspired by a large body of related work and principally advance the state of the art as follows:

\begin{enumerate}
\item rotor: generalization of symbolic execution through individual depth-first path exploration to simultaneous breadth-first exploration of control and data flow, and
\item bitme: bit-precise reasoning over RISC-V machine code through bounded model checking without high-level program analysis.
\end{enumerate}

The paper is structured as follows. We first discuss related work, then provide an overview of rotor, discuss principles, in particular soundness and completeness of rotor-generated models, followed by more details on how rotor works, has been validated for correctness, and may be used in practice. Next we introduce bitme and discuss its architecture and implementation. We then present our experimental results on rotor-generated models using bitme with Z3, bitwuzla, ADDs, and CFLOBDDs. We conclude with a discussion of future work.

\section{Related Work}

\subsection{Rotor}

Our tool \emph{rotor} constructs BTOR2~\cite{niemetz2018btor2} models of RISC-V~\cite{waterman2017risc} machines for RISC-V code synthesis and analysis for a \emph{bounded model checker}~\cite{niemetz2018btor2, biere1999symbolic} to explore. BTOR2 is related to Verilog~\cite{golze1996vlsi}, which is typically used as modeling language for hardware. Like BTOR2, Verilog allows modeling circuits (combinational logic) but only has limited support for modeling memory (sequential logic) such as registers. In contrast, BTOR2 allows us to capture the semantics of RISC-V through the theory of bitvectors and arrays of bitvectors, using the latter for modeling byte-addressed main memory.

Rotor builds a quantifier-free formula, linear in code size, in first-order logic for breadth-first exploration of control and data flow and targets SMT~\cite{brummayer2009boolector, niemetz2014boolector, niemetz2018btor2, de2008z3, niemetz2023bitwuzla} solvers to solve it (or domain-propagates it with bitme so that the SMT solver may, in fact, not be needed). This contrasts with \emph{symbolic execution engines}~\cite{cadar2008klee} that build individual formulae for each execution path (and thus do depth-first exploration). This way, rotor delegates the problem of explosion in control and data flow to SMT solvers. Checking individual paths in the model generated by rotor is left as an option in bitme.

Our motivation for breadth-first search is to study if SMT solvers and other bit-precise reasoning techniques can solve the entire problem. The only bounded model checker prior to this work that accepts BTOR2 models as input is BtorMC~\cite{niemetz2018btor2}, which is based on the Boolector~\cite{brummayer2009boolector} solver, the legacy predecessor of Bitwuzla~\cite{niemetz2023bitwuzla}.

\subsection{SMT Solvers}

The state-of-the-art \emph{SMT solvers} Z3~\cite{de2008z3} and Bitwuzla~\cite{niemetz2023bitwuzla} use \emph{bit-blasting} (combined with other approaches that we discuss in the next paragraphs). Bit-blasting is the process of translating (quantifier-free) higher-order logic SMT formulae to (quantifier-free) lower-order propositional logic formulae where all variables and constants have only Boolean values 0 or 1, and only Boolean operators are allowed. Those formulae are then solved by SAT solvers. It loses the high-level reasoning structure about integers and integer operations from the original SMT formula. Strategies using heuristics for solving those formulae are used.

Further, many SMT solvers, including Z3 and Bitwuzla, combine bit-blasting with other approaches to solving quantifier-free bitvector (QF\_BV) formulae. Bitwuzla implements \emph{propagation-based local search for satisfiability modulo theories}. It is based on word-level propagation of variable assignments, while also supporting propagation of constant bits~\cite{niemetz2020ternary}, similar to our work. But, unlike domain propagation in bitme, it starts down-propagation from the target (\emph{i.\,e.}, output) values and tries to find a satisfying assignment to input variables using word-level backtracking and randomization. It employs notions of \emph{inverse} and \emph{consistent} values that it tries to assign to inputs of an arithmetical operator or a logical operator in the expression tree at each propagation step. Experimental evaluation of propagation-based local search falling back to bit-blasting in Boolector showed up to 1000x speedup compared to only using bit-blasting~\cite{niemetz2016precise}. However, this combination still does not show good enough performance for our example models produced by rotor.

Z3 and Bitwuzla implement different approaches to support other theories on top of QF\_BV. Bitwuzla uses a lazy technique for incremental solving called \emph{lemmas on demand}~\cite{de2002lemmas, barrett2002checking} (supporting theory combinations as well) as its top-level framework and has theory solvers, one for each of the supported theories. Bitwuzla supports the (quantified and quantifier-free) theories of fixed-size bitvectors, arrays, floating-point arithmetic, and uninterpreted functions---where quantified formulae are handled by a quantifiers module~\cite{ge2009complete}, acting as a separate theory solver. (Z3~\cite{ge2009complete, wintersteiger2013efficiently} and Boolector~\cite{preiner2017counterexample} implement other techniques to support quantified bitvector formulae as well.) In Bitwuzla's solving engine, models are generated by the bitvector theory solver, and then, for the theory combination in the formula, respective theory solvers are used to check consistency with the theory axioms and produce refining lemmas. Our implementation of bitme currently supports the theory of arrays of bitvectors, translating it to the theory of bitvectors, as we describe in Section~\ref{sec:implementation}. Supporting other theories in bitme is future work.

Z3 uses another lazy technique CDCL($T$)~\cite{nieuwenhuis2006solving}---going down to SAT-level CDCL---to support theories. CDCL($T$) produces a skeleton for the output SAT formula to be solved using a CDCL-based SAT solver and uses a dedicated theory solver for each theory $T$ to refine the SAT solver with additional clauses it learns from the theory. CDCL first randomly assigns a value to a selected variable of the input SAT formula and then tries to infer values of the remaining variables to satisfy the formula. When it finds a conflict in the variable assignment, it adds a new clause to the formula. It then jumps to the level that lead to the conflict, and it sometimes restarts, keeping the memory of what it has learned from the previous conflicts. By repeating this way, it can find a satisfying variable assignment or find that the formula is unsatisfiable. Unlike local search, CDCL propagates assignments of variables in the direction from the input to the output. Our experiments show that domain propagation in bitme based on BDDs performs better than propagation in Z3 using CDCL($T$).

\subsection{Bitme}
\label{sec:bitme}

Currently existing model checkers suffer from slow performance, as they internally use SMT solvers, essentially translating their formulae to SAT. In this work, we explore using binary decision diagrams~\cite{bryant1986graph} (see Section~\ref{sec:bdds})---specifically, CFLOBDDs~\cite{sistla2024cflobdds, zhi2025polynomial}---in bounded model checking to reduce the number of queries to SMT solvers. We integrate our own implementation of CFLOBDDs, extended to a variant that we call CFLOBVDDs (Context-Free-Language Ordered Bit-Vector Decision Diagrams), into our model checking engine \emph{bitme}. CFLOBVDDs have a nested structure, like CFLOBDDs, but descend to BDDs with byte-sized bitvector input rather than bit-sized input. Bitme builds CFLOBVDDs for arithmetic and Boolean logic operators and the ternary (if-than-else) operator. Thus, it benefits from high-level reasoning about them, rather than going down to SAT.

Our experiments show that checking models generated by rotor using bitme---still partly falling back to using Z3 or Bitwuzla for certain examples---is faster than using Z3 or Bitwuzla alone, which largely rely on solving SAT problems. Also, one feature that could be seen as a benefit of bitme is that it can output all the found inputs that lead to an error, rather than one existing input, as done by local search or CDCL. Although, this is not a strong claim, since, in general, there can be exponentially many of such inputs.

While recent work shows that the performance of bit-blasting and CDCL($T$) can significantly benefit from combining local search with them~\cite{niemetz2016precise, zhang2024deep}, exploring a combination of local search and CFLOBDDs remains future work. Further, the paper by~\citet{lu2025btor2} proposes a machine-learning-based algorithm for selecting the most suitable model-checking technique for a given model. Integrating our tools with ML-based engines for finding the best-performing combination of approaches is another direction that remains to be explored in our future work.

\subsection{Binary Decision Diagrams}
\label{sec:bdds}

\emph{Binary decision diagrams} (BDDs)~\cite{bryant1986graph, clarke1997model} are used for bounded model checking. BDD is a way of representing a function table as a binary tree. It reasons about input to the program on bit level. For the first input bit, which becomes the root node of the tree, it has two branches for its possible values, 0 and 1. Each of the two children of the root node can be either a leaf node, whose value is, as well, 0 or 1, or an intermediate node---a subtree for the next input bit. This way, it decides on program output depending on input. The downside is that the depth of the tree, which can potentially exponentially explode, highly depends on the choice of ordering the input bits. Heuristics exist for choosing a more optimal order, but, in general, finding such a choice itself may take exponential time.

\emph{Algebraic decision diagrams} (ADDs)~\cite{frohm1993algebraic} are extended BDDs whose output nodes' values are integers rather than Boolean constants. In bitme, we generalize ADDs so that they have byte-sized, rather than bit-sized, input nodes. The consequence is that each non-leaf node of such an ADD has 256 child nodes, unlike 2 child nodes in standard ADDs or BDDs. Thus, they are represented by trees that have lesser depth at the cost of greater width at each tree level, as compared to classical BDDs with bit-sized input.

\emph{Context-Free-Language Ordered Binary Decision Diagrams}~\cite{sistla2024cflobdds, zhi2025polynomial} (CFLOBDDs) are a recent extension to the BDD line of work. The key ideas are, firstly, to express the decision paths of the model, very similar to execution paths in a computer program, and, secondly, to reuse parts of the decision tree rather than building recurring parts doing the same computations. The decision-making parts of a CFLOBDD are nested smaller CFLOBDDs. Specifically, a CFLOBDD of level $l$ consists of CFLOBDDs of level $l - 1$, going down to level $0$, which corresponds to standard BDDs.

Recent work on bounded model checking~\cite{biere1999symbolic, niemetz2018btor2} has moved from using classical BDDs to other approaches, because BDDs may blow up in space for more complicated programs and thus may not scale well. We have chosen to implement the CFLOBVDD (see Section~\ref{sec:bitme}) variant of BDDs in bitme, since CFLOBDDs, and thus CFLOBVDDs, are designed to allow reusing parts of decision trees due to their nested structure, and so may scale, for the best-case ordering of inputs, exponentially better. Here, the reasoning for extending CFLOBDDs to CFLOBVDDs is the same as the reasoning for extending ADDs above. We choose CFLOBDDs over other BDD variants, such as, for example, FBDDs~\cite{wegener2000branching}, BMDs~\cite{bryant1995verification}, EVBDDs~\cite{lai1992edge}, or FEVBDDs~\cite{tafertshofer1997factored}, because exponential compression found in other BDDs carries over to CFLOBDDs~\cite{wegener2000branching, bryant1995verification, clarke1995hybrid, jain1997indexed} without weights that may involve an unbounded number of bits~\cite{lai1992edge, tafertshofer1997factored}.

\section{Model Generation with Rotor}

We first provide an overview of rotor, then discuss principles, in particular soundness and completeness of rotor-generated models, followed by more details on how rotor works, has been validated for correctness, and may be used in practice.

\begin{figure}
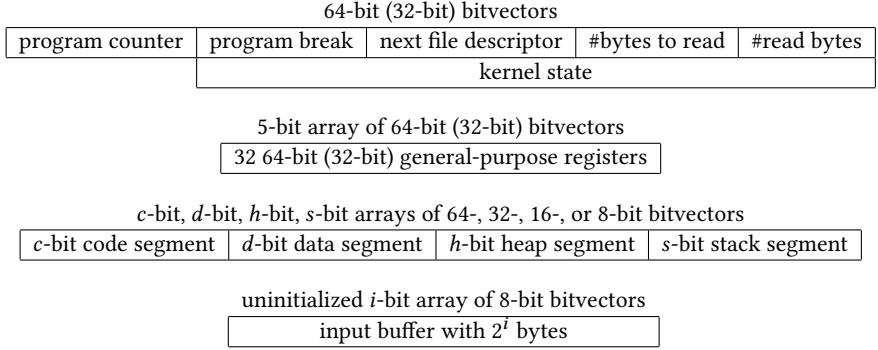

\centering
\begin{tabular}{|c|c|c|c|c|}
\multicolumn{5}{c}{64-bit (32-bit) bitvectors} \\
\hline
program counter & program break & next file descriptor & \#bytes to read & \#read bytes \\
\hline
\multicolumn{1}{c}{} & \multicolumn{4}{|c|}{kernel state} \\
\cline{2-5}
\end{tabular}
\vspace{1em}\\
\begin{tabular}{|c|}
\multicolumn{1}{c}{5-bit array of 64-bit (32-bit) bitvectors} \\
\hline
32 64-bit (32-bit) general-purpose registers \\
\hline
\end{tabular}
\vspace{1em}\\
\begin{tabular}{|c|c|c|c|}
\multicolumn{4}{c}{$c$-bit, $d$-bit, $h$-bit, $s$-bit arrays of 64-, 32-, 16-, or 8-bit bitvectors} \\
\hline
$c$-bit code segment & $d$-bit data segment & $h$-bit heap segment & $s$-bit stack segment \\
\hline
\end{tabular}
\vspace{1em}\\
\begin{tabular}{|c|}
\multicolumn{1}{c}{uninitialized $i$-bit array of 8-bit bitvectors} \\
\hline
input buffer with $2^i$ bytes \\
\hline
\end{tabular}
\caption{\label{fig:state}64-bit (32-bit) RISC-V machine state: each bitvector and array of bitvectors is updated by a separate transition function in pure combinational BTOR2 logic. All bitvectors other than those of the input buffer are initialized with constants. Size of model input is bounded by $2^i$ bytes, \emph{i.\,e}., each byte is only read at most once.}
\end{figure}

Rotor is a tool for generating bit-precise user-space 64-bit and 32-bit RISC-V machine models using combinational and sequential logic over bitvectors (program counter, kernel state) and arrays of bitvectors (register file, memory, I/O) represented in BTOR2~\cite{niemetz2018btor2}, see Fig.~\ref{fig:state}. Reasoning about rotor-generated models involves bounded model checking, either externally using a bounded model checker such as BtorMC~\cite{niemetz2018btor2} or bitme as introduced here, or internally by removing sequential logic from models in rotor to target SMT solvers such as Z3 or Bitwuzla. For this purpose, rotor may also generate models without sequential logic in SMT-LIB~\cite{barrett2016satisfiability}. BTOR2 without sequential logic is essentially equivalent in semantics to SMT-LIB over bitvectors and arrays of bitvectors.

Rotor supports RISC-V with integer arithmetic only, that is, around 100 machine instructions. In particular, rotor supports 64-bit and 32-bit integer arithmetic (RV64I, RV32I extensions) with multiplication and division (RV64M, RV32M) as well as compressed instructions (RVC). Support of floating-point arithmetic and other extensions is future work.

Rotor models user-space 64-bit (32-bit) RISC-V machines with a 64-bit (32-bit) program counter, 32 64-bit (32-bit) general-purpose registers and a 32-bit address space, that is, 4GB of byte-addressed main memory that is segmented into code, data, heap, and stack segments. Instead of modeling priviledged instructions, we model the functionality of selected system calls for termination (\texttt{exit}), heap allocation (\texttt{brk}), and I/O (\texttt{openat}, \texttt{read}, \texttt{write}) according to RISC-V conventions.

The modeled kernel state involves 64-bit (32-bit) bitvectors for the program break (\texttt{brk}), the next file descriptor (\texttt{openat}), and two counters for keeping track of how many bytes have already been read in a currently executing \texttt{read} system call as well as in total. Finally, there is an (uninitialized) array of 8-bit bitvectors that represents an input buffer of bytes to be read by the \texttt{read} system call. The \texttt{read} system call is modeled to fail after all bytes have been read. Support of other failure modes is future work. In particular, the \texttt{write} system call is modeled to succeed without doing anything else. Side effects outside of models through I/O are ignored.

Main memory is modeled through segmentation into code, data, heap, and stack segments. Each segment is modeled by an array of bitvectors that can be configured to be 64-bit, 32-bit, 16-bit, or 8-bit bitvectors. Segmentation allows rotor to control array size but requires modeling virtual-to-physical address translation. Segmentation is common in tools related to rotor, such as KLEE~\cite{cadar2008klee}. Program counter, registers, and segments are initialized according to RISC-V conventions. In particular, the stack segment is initialized with console arguments of executables. However, support of console arguments as symbolic input is future work.

\paragraph{Safety Properties.}

Rotor generates checks for safety properties such as invalid machine instructions and memory addresses, unknown machine instructions and system call IDs, non-zero exit codes, division by zero, signed division overflow, and segmentation faults caused when fetching code, loading and storing, and executing system calls. Checks may be omitted to increase solver performance. Support of other checks such as memory safety and even liveness checks is possible but remains future work.

\paragraph{Code Synthesis.}

Invoking "\texttt{rotor - 1}" generates a 64-bit RISC-V machine model (around 200KB) with an uninitialized code segment and all safety checks encoded as constraints except for the non-zero exit code check ("\texttt{rotor -m32 - 1}" generates a 32-bit version that is around half the size of the 64-bit version). Bounded-model checking such a model corresponds to code synthesis by exploring code segment content that terminates with exit code \texttt{1} when executed. This does indeed work and results in executable code that resembles a proper \texttt{exit} system call wrapper implemented in RISC-V machine code. We have nevertheless not yet explored code synthesis further.

\paragraph{Code Analysis.}

With an initialized code segment, the input buffer for \texttt{read} system calls is the only source of symbolic non-constant input to rotor-generated models. Invoking "\texttt{rotor -l exec - 1}" generates a RISC-V machine model initialized with the content of an executable ELF RISC-V binary \texttt{exec}. The result is a 64-bit or 32-bit model depending on whether \texttt{exec} is a 64-bit or 32-bit binary, respectively. Bounded-model checking such a model corresponds to code analysis by exploring machine inputs on which \texttt{exec} terminates with exit code \texttt{1} when executed.

\paragraph{Program Equivalence.}

Rotor supports an experimental feature for generating RISC-V machine models with multiple cores and shared, partially shared, and unshared memory that enable checking program equivalence for code analysis but also for code synthesis. For example, rotor may generate a two-core model with unshared memory initialized with two different executables. Bounded-model checking such a model corresponds to code analysis by exploring machine inputs on which both executables terminate, say, with the same exit code, when executed, possibly involving differently many executed machine instructions. Checking program equivalence for all machine inputs represents an interesting use case for universal quantification of the input buffer. Alternatively, the code segment of one of the cores may remain uninitialized. In this case, bounded model checking again corresponds to code synthesis, representing an interesting use case for mixed universal (input buffer) and existential (code segment) quantification. We have nevertheless not yet explored checking program equivalence further. Support of multithreaded code may be possible as well but also remains future work. Our focus here is on checking safety properties of RISC-V machine code.

\subsection{Principles}

Next we mildly formalize rotor to establish claims on soundness and completeness. Formal verification of those claims is a significant challenge that is orthogonal to the goals of this paper and remains future work. However, we have developed tests that helped us debug parts of model generation, as discussed further below.

\begin{definition}
\label{def:rotor}
A \emph{rotor-generated RISC-V machine model} $R$ is a tuple $(S, I, T, B, C)$ where $S$ is a set of machine states, $I\subseteq S$ is a set of initial states, $T:S\rightarrow S$ is a transition function, $B\subseteq S$ is a set of \emph{bad} machine states, and $C\subseteq S$ is a set of (good) \emph{constraint} machine states with $B\cap C=\emptyset$.
\end{definition}

A machine state is a set of instances of a given set of bitvectors $V$ and arrays of bitvectors $A$. The set $S$ of machine states is declared by BTOR2 \texttt{state} operators, one for each bitvector and array of bitvectors. An \emph{instance} of an $n$-bit bitvector $v$ is a pair $(v,c)$ where $c$ is an $n$-bit binary value of $v$ out of the $2^n$ possible values of $v$, as specified by a BTOR2 \texttt{const} operator. An instance of an $m$-bit array $a$ of $n$-bit bitvectors is a set of pairs $((a,i),c_i)$ for all $m$-bit binary values $0\leq i<2^m$ of an $m$-bit (index) bitvector where $c_i$ is an $n$-bit binary value of an $n$-bit (element) bitvector.

For all bitvectors $v\in V$, there is either the same instance of $v$ in $s$ for all $s\in I$, as specified by the BTOR2 \texttt{init} operator, or else, if there is an initial state $s\in I$ with an instance of $v$, then there are initial states $t\in I$ for all instances of $v$ where $t$ is otherwise equivalent to $s$. The latter case refers to uninitialized bitvectors which are specified as such by the absence of a BTOR2 \texttt{init} operator. The same holds analogously for arrays of bitvectors in initial states. The specification in BTOR2 is more involved and skipped here. Machine input, and even to-be-synthesized machine code, is given through uninitialized arrays of bitvectors denoted $MI$ and $MC$, respectively.

The transition function $T$ is a total function from machine states to machine states where $T$ is represented by a set of functions $T_v:S\rightarrow v$ and $T_a:S\rightarrow a$ for each bitvector $v\in V$ and array of bitvectors $a\in A$ in $S$, respectively. Each function $T_v$ and $T_a$ is a bitvector and array-of-bitvector combinational-logic BTOR2 or SMT-LIB expression, connected to $v$ and $a$, respectively, with a BTOR2 \texttt{next} operator, which is the only true sequential-logic operator in BTOR2.

The set of \emph{bad} and (good) \emph{constraint} machine states represent proof obligations, also referred to as bad and good properties, as specified by Boolean combinational-logic BTOR2 or SMT-LIB expressions over subsets of the bitvectors $V$ and arrays of bitvectors $A$, and further distinguished by the BTOR2 \texttt{bad} and \texttt{constraint} operators, respectively.

\begin{definition}
Given a natural number $k\geq 0$, the $k$-\emph{transition} function of a rotor-generated RISC-V machine model $(S, I, T, B, C)$ is a function $T_k:S\rightarrow S$ where $T_0(s)=s$ and $T_i(s)=T(T_{i-1}(s))$ for all $0<i\leq k$ and $s\in S$.
\end{definition}

In our context, bounded model checking computes $T_k$ for increasing $k$ and checks, for all $s\in I$ and each $k$, the satisfiability of all bad properties applied to $T_k(s)$ constrained to all good properties, also referred to as the $k$-\emph{satisfiability} of a model. Applying the transition function $T$ to a machine state is what constitutes a \emph{transition}. The process of computing $T_k$ is referred to as model \emph{unrolling} while $T_k(s)$ is called the $k$-\emph{transition} of a machine state $s\in S$. Unrolling essentially removes sequential logic, that is, state from models turning the solver problem into a pure combinational-logic problem.

\begin{definition}
A rotor-generated RISC-V machine model $(S, I, T, B, C)$ is $k$-\emph{satisfiable} with $k\geq 0$ for an initial state $s_0\in I$ if $T_k(s_0)\in B$ and $T_i(s_0)\in C$ for all $0\leq i\leq k$.
\end{definition}

We say that a model is $k$-\emph{unsatisfiable} if the model is not $k$-satisfiable. This may happen if $T_k(s_0)\notin B$ for all $s_0\in I$, that is, the $k$-transitions of all initial states are not bad machines states. It may also happen if for all $s_0\in I$ with $T_k(s_0)\in B$, there is an $0\leq i\leq k$ such that $T_i(s_0)\notin C$, that is, there are $i$-transitions of all initial states whose $k$-transitions are bad machine states that are not (good) contraint machine states. We may also be interested in the least $k$ for which a model is $k$-satisfiable.

\begin{definition}
A rotor-generated RISC-V machine model $(S, I, T, B, C)$ is \emph{least}-$k$-\emph{satisfiable} with $k\geq 0$ if the model is $k$-satisfiable and $T_i(s_0)\notin B$ for all initial states $s_0\in I$ and $0\leq i<k$.
\end{definition}

A rotor-generated model may be $k$-satisfiable (least-$k$-satisfiable) for more than one initial state if there are uninitialized bitvectors or arrays of bitvectors. We say that a model is $k$-satisfiable for machine input $IN$ and machine code $CO$ if the model is $k$-satisfiable (least-$k$-satisfiable) for an initial state with instances $(MI,IN)$ and $(MC,CO)$. Rotor-generated models are designed to be sound and complete in the following sense.

\begin{proposition}
\label{prop:rotor}
Given a rotor-generated RISC-V machine model $(S, I, T, B, C)$, for all $k\geq 0$, a machine state $s\in B$ is \emph{reachable} on a RISC-V machine by executing no more than (at least) $k$ machine instructions of machine code $CO$ on machine input $IN$ if and only if the model is $k$-satisfiable (least-$k$-satisfiable) for machine input $IN$ and machine code $CO$.
\end{proposition}

The proposition holds if \texttt{read} system calls only consume one input byte per invocation. A \texttt{read} system call that consumes $n$ input bytes with $n>1$ must be accounted for as $n$rather than just one machine instruction. Generalizing the proposition accordingly is straightforward but skipped here for simplicity.

Rotor reduces the problem of determining reachability of RISC-V machine states to the problem of determining satisfiability of BTOR2 models. Formally proving the proposition is a major challenge and remains future work. Finding $k$ for which a model is $k$-satisfiable through unrolling is the job of a bounded model checker such as btormc or bitme. However, rotor can also simulate bounded model checking by unrolling models itself.

\begin{definition}
Given some $k\geq 0$ and a rotor-generated RISC-V machine model $(S, I, T, B, C)$, its $k$-\emph{unrolled} model is the model $(S, I, T_k, B, C)$ where $T_k$ is the $k$-transition function of $T$.
\end{definition}

The transition function of a $k$-unrolled model is specified by pure bitvector and array-of-bitvector combinational-logic BTOR2 and SMT-LIB expressions without any sequential logic. Hence $k$-unrolled models can be solved by SMT solvers without a bounded model checker.

\begin{corollary}
Given a rotor-generated RISC-V machine model $R$, for all $k\geq 0$, the $k$-unrolled model of $R$ is $0$-satisfiable if and only if $R$ is $k$-satisfiable.
\end{corollary}

Given some $k\geq 0$, rotor can unroll models that also include all $i$-unrolled models with $0\leq i\leq k$, for the purpose of checking least-$k$-satisfiability.

\subsection{Implementation}
\label{sec:implementation}

Rotor is implemented in around 14k lines of code written in C* which is a tiny subset of C originally developed for teaching~\cite{Onward17}. Rotor is based on the selfie system~\cite{Onward17}, written in 12k lines of C* code, bringing the total code base of rotor as a system to around 26k lines of code. The code is open source under a two-clause BSD license.

\paragraph{History.}

The rationale behind rotor is probably best understood by taking a quick look at its development, including the publicly available history of the selfie system. Development of selfie apparently began around 2015 while our work on rotor started in late 2023. The original purpose of selfie is said to be educational for teaching undergraduate compiler and systems classes. Selfie implements in a single self-contained C* file (without any imports): a self-compiling compiler that targets RISC-U, which is a 14-instruction subset of 64-bit and 32-bit RISC-V limited to unsigned integer arithmetic; a self-executing RISC-U emulator; and a self-hosting RISC-U hypervisor~\cite{Onward17}. Selfie compiles itself as well as executes and virtualizes itself, hence the name~\cite{Onward17}. Selfie rests on the assumption that understanding self-reference as it appears in programming languages, compilers, and systems is one of the key challenges in computer science education~\cite{Onward17}.

\paragraph{Vision.}

The original idea behind rotor is to push that vision further by introducing formal methods into undergraduate teaching. Rotor essentially provides an \emph{inverted} perspective on semantics based on mathematical logic, enabling students to see the \emph{inverse} of what computers actually do. Solving rotor-generated models is essentially computing the inverse of what a RISC-V machine does, including a machine running rotor. Inspired by selfie, rotor is able to model itself (its BTOR2 model is around 28MB, dominated by model initialization which is linear in the size of the around 500KB rotor executable). Principles and simplicity has been key to bootstrapping rotor. However, there are limits, of course. Turns out that even state-of-the-art SMT solvers struggle with rotor-generated models, as shown in our experiments, motivating us to develop bitme for taking a closer look.

\paragraph{Model Generation.}

Rotor generates RISC-V machine models in constant time and space. Given a RISC-V executable, model generation and size go up to linear in the size of the executable. In short, rotor is fast and scales. Rotor may load executables generated with gcc or the selfie compiler which produces unoptimized RISC-U machine code. Rotor may also compile C* files using the selfie compiler internally, effectively connecting model generation to C* via RISC-U, and even C or any other language that compiles to RISC-V with integer arithmetic. This means that rotor can be understood as benchmarking tool that enables studying and eventually improving SMT solver performance when reasoning about programs written in high-level programming languages.

\paragraph{Model Structure.}

Rotor generates models by declaring and initializing bitvectors and arrays of bitvectors that represent machine states first. Next, transition functions for updating each of those bitvectors and arrays are generated constituting the bulk of combinational logic in a model. Lastly, Boolean expressions in combinational logic representing bad and good properties are generated. Each expression may refer to the current machine state by plugging into the bitvectors and arrays representing state, and the next machine state by plugging into transition functions.

\paragraph{Common Subexpressions.}

Transition functions as well as bad and good properties are all pure functions. We therefore wrote the code that generates those functions following a programming style that resembles functional programming in C* with one notable disclaimer. Common subexpressions are in fact very common in rotor-generated models. For example, the RISC-V decoder logic is used as many times as there are RISC-V machine instructions in a model. In order to reduce the number of duplicates, we implemented a naive, as in syntactical, equivalence check for reusing previously generated operations that share the same operands with operations that need to be generated again. Reusing subexpressions involves searching previously generated subexpressions which we keep in a singly-linked list. Fortunately, model size without executable code is constant and relatively small, making the otherwise square complexity of reusing common subexpressions this way negligible. However, it does impact performance when generating model initialization for executable code, as this is linear in code size, which is why we turn it off for that. Here, the increase in number of duplicates is bounded by a small constant and therefore again negligible, allowing us to keep the code for reusing subexpressions simple.

A RISC-V machine, like any von Neumann architecture, works in a simple cycle: fetch, decode, execute, and so on. That cycle is reflected in rotor-generated models as follows.

\paragraph{Fetch.}

In order to support uncompressed as well as compressed RISC-V machine instructions, enabling gcc support in particular, there are two functions that model fetching of code, one for uncompressed instructions encoded in 32 bits and one for compressed instructions encoded in 16 bits. In the presence of compressed instructions, uncompressed instructions are only guaranteed to be 16-bit-aligned but not 32-bit-aligned in the code segment, making fetching code a bit more complicated. Fetching code involves the bitvector that represents the program counter and the array that represents the code segment whose content is constant (or uninitialized for code synthesis). During bounded model checking the value of the program counter becomes symbolic as soon as control flow depends on machine input.

\paragraph{Decode.}

Decoding of code involves a number of functions corresponding to a full RISC-V decoder which are therefore quite complex. Yet these functions only connect to the output of the functions that model fetching of code. Essentially, there is a decoding function for determining which machine instruction is currently present and decoding functions for determining each of the possible arguments. We decided to model decoding of compressed instructions separately, featuring separate execution logic for compressed instructions. There is an alternative though since compressed instructions beautifully map to uncompressed instructions in semantics, one of many features that make RISC-V a major achievement. Hence we could model decoding of compressed instructions by mapping them to uncompressed instructions first before decoding uncompressed instructions. However, we chose not to do so as the execution logic for compressed instructions is considerably simpler than the execution logic for uncompressed instructions. Since compressed instructions occur more frequently in gcc-generated code that decision may eventually pay off in better solver performance. Substantiating that in experiments remains future work.

\paragraph{Execute.}

Fortunately, the size of side effects on machine state when executing RISC-V code is strictly bounded by at most 128 bits (64 bits on 32-bit machines) that may change when executing a single RISC-V machine instruction, with the notable exception of \texttt{read} system calls. Otherwise, only the 64-bit (32-bit) value of the program counter may change and either the value of a single 64-bit (32-bit) general-purpose register or else the value of a 64-bit (32-bit) machine word in main memory. This is another feature of RISC-V that simplifies modeling. Only a \texttt{read} system call that consumes $n$ input bytes may, in addition to the value of the program counter and one register carrying the return value of the call, change the values of up to $n$ bytes in main memory where the buffer for storing input bytes is located. In order to establish a still constant 128+8-bit (64+8-bit) bound on the size of per-transition changes in machine state in rotor-generated models (without kernel state), we model \texttt{read} system calls by preventing the program counter from advancing until the $n$ input bytes have been consumed and stored in memory, one byte at a time per transition.

Execution logic is modeled by transition functions for control flow updating the bitvector that models the program counter, register data flow updating the array that models the register file, and memory data flow updating the arrays that model the data, heap, and stack segments. Another key feature of RISC-V that simplifies modeling here is that the semantics of all RISC-V machine instructions matches the standard bitvector semantics of SMT solvers. However, there still remains the challenge of handling kernel activity through system calls affecting control and data flow in registers and memory. Fortunately, the kernel can only be invoked by a single machine instruction with no arguments called an \texttt{ecall} instruction. Arguments to the kernel are assumed to be stored in dedicated registers per RISC-V conventions.

\paragraph{Kernel State.}

There are four transition functions for the program break maintained by the \texttt{brk} system call, the next file descriptor returned by the \texttt{openat} system call, and the two counters keeping track of the number of input bytes still-to-be read by a \texttt{read} system call and the input bytes already read in total. Segmentation faults caused by system calls are tracked by safety properties. Upon detecting an \texttt{ecall} instruction in the decoder, the invoked system call and its arguments are determined by reading from specific registers per RISC-V conventions. Since the register file is modeled by an array of bitvectors, individual registers are read by the BTOR2 \texttt{read} operator. Modeling of the \texttt{exit} system call is unique as it is the only machine activity that results in all subsequent transitions having no effect on machine state, in particular the program counter, signalling that the machine has terminated execution which in turn can be detected by a bounded model checker.

\paragraph{Control Flow.}

The transition function for updating the program counter models the control-flow semantics of all machine instructions, uncompressed and compressed, as well as all system calls. The function is essentially a large if-then-else expression over all possible choices provided by the decoder and the logic for identifying system calls. Only \texttt{exit} and \texttt{read} system calls involve multi-transition control-flow behavior which requires the program counter not to advance, either at all, with the former, or else, with the latter, until the input bytes that are supposed to and can be read have actually been read into memory. Register-relative and conditional control flow also involves reading from the program counter and the registers.

\paragraph{Register Data Flow.}

The transition function for updating the register file models the register data-flow semantics of all machine instructions that feature register data flow as well as all system calls but the \texttt{exit} system call, as system calls update a dedicated register for their return value per RISC-V conventions. Since the register file is modeled by an array of bitvectors, the transition function is a function that updates an array, and not just a bitvector, using the BTOR2 \texttt{write} operator, in addition to the BTOR2 \texttt{read} operator for reading from individual registers. Yet only a single bitvector in the array may be updated per transition as only a single register may be updated when executing RISC-V machine code including system calls. Since RISC-V \texttt{load} instructions read from memory, the transition function may read from the arrays of bitvectors that model the data, heap, and stack segments. The \texttt{openat} and \texttt{write} system calls do that in reality too but not in our model. Their unmodeled memory access is nevertheless checked for segmentation faults. Reading from memory, as well as writing to memory as discussed next, involves translating virtual memory addresses to physical segment addresses. The translation is non-trivial as memory access may not be memory-word-aligned in segments, generally requiring multiple BTOR2 \texttt{read} operations followed by combining multiple partial memory words into proper machine words. There are safety properties for checking segmentation faults for read access, and separately for write access as discussed next.

\paragraph{Memory Data Flow.}

There are three transition functions for updating the data, heap, and stack segments modeling the memory data-flow semantics of RISC-V \texttt{store} instructions as well as the \texttt{read} system call which writes input bytes to memory but only in the heap segment. Reading from registers is again facilitated by the BTOR2 \texttt{read} operator. Similar to reading from memory, writing to memory involves address translation, and in general, slicing machine words into multiple memory words which in turn are written to segments using multiple BTOR2 \texttt{write} operations. All memory stores are checked for segmentation faults in safety properties as mentioned before.

\paragraph{Modeling versus Reasoning.}

There are essentially three sources of complexity in modeling that may have an immediate impact on model checking performance: safety properties, that is, the number and type of properties to be checked; virtual memory, that is, large arrays of bitvectors, in particular the access logic for the arrays modeling memory segments; and instruction handling, that is, the fetch and decoder functions as well as the execution logic of RISC-V instructions. By default rotor includes in generated models all supported safety properties such as illegal instructions and addresses, non-zero exit codes, division by zero, and segmentation faults. For better model checking performance some of those checks may be turned off.

For virtual memory modeling, virtual address spaces may be configured to be smaller or bigger than 32 bits, and memory words may be configured to be bytes, 16-bit half words, 32-bit single words, or 64-bit double words. The bigger the memory words are the smaller the address spaces of segments, that is, the array index spaces are, and the wider the memory bus is between CPU and memory. However, chances of complex unaligned memory access also increase. We present experimental data that demonstrates the effects of different configurations on model checking performance.

Instruction handling may be optimized and even traded off between model generation and model checking. Rotor can be configured to turn off support of certain RISC-V extensions such as compressed instructions and even any RISC-V instructions that are not RISC-U instructions, considerably decreasing the size and complexity of the decoder and execution logic. A tool preceding rotor called \emph{beator} developed by us went even further but has been abandoned with rotor. The idea is to model control and data flow of RISC-V machine code directly in BTOR2 where all instruction fetching and decoding is done in beator during model generation. Hence code synthesis is not supported with beator-generated models. Moreover, the size of the execution logic in beator-generated models is not constant but linear in the size of the modeled executable. In contrast, rotor utilizes an array of bitvectors initialized by constants that models the code segment, keeping the size of the execution logic constant. Either way, beator versus rotor hints at a principled tradeoff when to fetch and decode, at model generation or model checking time, or possibly even through some form of controlled pre-fetching and pre-decoding during model checking that may deserve to be explored in future work.

\subsection{Correctness.}

The code that implements rotor is highly complex, not in terms of algorithmic complexity, but in terms of low-level bit-precise manipulation of data. During development we heavily relied on bounded model checkers such as btormc and later bitme for testing soundness and completeness of models generated by rotor for handwritten C code with known safety properties. The RISC-U emulator in rotor also played an important role in determining and validating machine input and the exact number of executed instructions when safety properties are violated during code execution. There is, however, an additional measure that we took, not just for debugging but also for performance and even for its educational value when using rotor in class.

\paragraph{BTOR2 Semantics.}

We designed and implemented a BTOR2 emulator, that is, an emulator for evaluating BTOR2 expressions through constant propagation, showcasing BTOR2 semantics on constants. The emulator can even be used on models with symbolic input but only naively by evaluating a given model on all input values enumerated as constants. This only works with models that read no more than a single byte. Hence rotor may indeed evaluate some models it generates and thereby predict what bounded model checkers and SMT solvers should report on those models. For example, rotor is able to determine machine input as well as lower and upper bounds on the number of model transitions when safety properties are violated.

\paragraph{Disassembler.}

Since rotor-generated models contain a RISC-V decoder, we integrated a RISC-V disassembler into the BTOR2 emulator which outputs RISC-V assembly during model evaluation, on enumerated machine input for debugging models, and by iterating over the code segment of models to generate RISC-V assembly for a given executable. We compared the RISC-V assembly generated by rotor with the RISC-V assembly generated by the official RISC-V toolchain which helped us identify and fix bugs in rotor.

\paragraph{Constant Propagation.}

When unrolling models into pure combinational logic, rotor may utilize the BTOR2 emulator for constant propagation to remove all BTOR2 expressions from the unrolled model that do not depend on machine input. The effect is significant in terms of model size while the impact on solver performance is generally noticable. Constant propagation is also done in bitme where it also has an impact on solver performance, especially when combined with other measures.

\subsection{Usage}

\begin{table}
\centering
\begin{tabular}{|l|l|}
\hline
\textbf{console argument} & \textbf{description}\\
\hline\hline
\multicolumn{2}{|c|}{(selected) selfie frontend arguments}\\
\hline
\texttt{-c \{C\_file\_name\}} & generate linked RISC-U code for C* files\\
\texttt{-l executable\_name} & load selfie- or gcc-generated ELF executable\\
\texttt{-m64} & generate 64-bit RISC-V machine code and models (default)\\
\texttt{-m32} & generate 32-bit RISC-V machine code and models\\
\hline\hline
\multicolumn{2}{|c|}{(selected) rotor backend arguments}\\
\hline
\texttt{-o model\_name} & specify name of generated model (default: frontend C file name)\\
\texttt{-m} & evaluate model through constant propagation\\
\texttt{-d} & output RISC-V assembly during model evaluation\\
\texttt{-s} & generate RISC-V assembly by applying model to loaded code\\
\texttt{-l} & load additional code into separate core of multicore model\\
\hline
\texttt{-Pnobadexitcode} & do not include check for bad exit code\\
\texttt{-Pgoodexitcode} & include check for good exit code\\
\texttt{-Pnoexitcodes} & do not include check for exit codes\\
\texttt{-Pnodivisionbyzero} & do not include check for division by zero\\
\texttt{-Pnodivisionoverflow} & do not include check for division overflow\\
\texttt{-Pnoinvalidaddresses} & do not include check for invalid addresses\\
\texttt{-Pnosegfaults} & do not include check for segmentation faults\\
\hline
\texttt{-bytestoread b} & model \texttt{read} system call to fail after reading $b$ bytes (default: 1)\\
\texttt{-cores c} & generate $c$ cores (default: 1)\\
\texttt{-virtualaddressspace a} & set virtual address space to $a$ bits (default: 32)\\
\texttt{-codewordsize w} & set code word size to $w$ bits (default: 32)\\
\texttt{-memorywordsize w} & set memory word size to $w$ bits (default: 64)\\
\texttt{-heapallowance a} & set heap allowance to $a$ bytes (default: 4096)\\
\texttt{-stackallowance a} & set stack allowance to $a$ bytes (default: 2048)\\
\hline
\texttt{-riscuonly} & generate RISC-U machine models\\
\texttt{-noRVC} & omit model of compressed instructions\\
\texttt{-noRVM} & omit model of multiplication and division instructions\\
\hline
\texttt{-kmin k} & unroll model for k steps skipping bad machine states up to k\\
\texttt{-kmax k} & unroll model for k steps\\
\texttt{-k} & determine $k_{min}$ and $k_{max}$ for single-byte-input models\\
\texttt{-sat} & only include satisfiable bad machine states in -k unrolled models\\
\texttt{-smt} & generate unrolled SMT-LIB model\\
\hline
\end{tabular}
\caption{\texttt{rotor \{frontend arguments\} - exit\_code \{backend arguments\}}}
\label{tab:rotor-options}
\end{table}

Table~\ref{tab:rotor-options} provides a summary of most rotor console arguments which have been described above. We distinguish rotor frontend and backend arguments. Frontend arguments control the builtin selfie C* compiler and RISC-V bootloader. The backend arguments control model generation and evaluation as previously mentioned. The arguments for controlling model generation with multiple RISC-V cores are not used in our experiments.

\section{Bounded Model Checking with Bitme}

Similar to rotor, we first provide an overview of bitme, then discuss principles, in particular soundness and completeness of bitme, followed by more details on how bitme works, has been validated for correctness, and may be used in practice.

Bitme is a tool for bounded model checking rotor-generated BTOR2 models. Bitme supports a subset of BTOR2. In particular, support of BTOR2 operators for specifying fairness and liveness properties~\cite{niemetz2018btor2} is future work.

Bitme essentially unrolls rotor-generated BTOR2 models into pure combinational-logic expressions and utilizes the Z3 and bitwuzla SMT solvers for checking satisfiability of those expressions. Bitme also features implementations of binary decision diagrams (BDDs)~\cite{bryant1986graph}, namely algebraic decision diagrams (ADDs)~\cite{frohm1993algebraic} and context-free-language ordered binary decision diagrams (CFLOBDDs)~\cite{sistla2024cflobdds} that we slightly generalized to integrate with model input that is provided in bytes and not bits. In bitme, BDDs map model input to model state to determine satisfiability without using SMT solvers at all or at least deferring their use.

Given a rotor-generated model, bitme first initializes the model and asserts all good properties on the initialized model. Bitme reports any good but unsatisfiable properties and terminates if there are any. Otherwise, bitme moves on and checks satisfiability of all bad properties and reports satisfiable bad properties along with all or some satisfying model input. Then, bitme may optionally assert negations of all bad properties. After that, bitme unrolls the model for the next transition by asserting all transition functions applied to the current state of the model as equivalent to the next state. Optionally, if no state change is detected, bitme reports that and terminates. Otherwise, bitme completes the next transition, either by looping back to asserting all good properties on the next state and so on, or by terminating if the bound on model unrollings has been reached. Optionally, bitme may explore control flow depth-first at this point by recursively asserting rotor-generated conditions on control flow, first by asserting that control flow branches before looping back, and then, upon returning from the recursion, by asserting that control flow does not branch before looping back again.

In order to defer and even avoid the use of SMT solvers, bitme uses BDDs to implement \emph{domain propagation}~\cite{schulte2005bounds}. The idea is here to propagate the domain of variables through models by enumerating all possible values of a variable and then propagating those values as constants while keeping track of the original values during propagation. Variables are here bitvectors that are uninitialized in a model such as the bitvectors that represent model input. Domain propagation involves a \emph{decision tree} that we call an (input-to-state) \emph{tracker} mapping input values to state values, that is, model input to model state during unrolling. In general, an input value can only be mapped to a single state value. However, different input values may be mapped to the same state value. In short, a tracker is a total function that can represent or \emph{track} any combinational-logic expression, in particular $k$-transition functions and safety properties over those functions.

\subsection{Principles}

\begin{definition}
Given a combinational-logic expression $e$, a \emph{tracker} of $e$ is a total function $t:I\rightarrow S$ from the input values $I$ to the output (state) values $S$ of $e$ such that $t(i)=e(i)$ for all $i\in I$.
\end{definition}

In bitme, domain propagation is limited to domains containing no more than 256 values, that is, bitvectors of up to 8 bits in size. While the size of domains is limited, the size of codomains is not since only the size of the actual range of a tracker matters in practice, that is, the size of model states at any point during model unrolling. However, bitme only supports codomains representing bitvectors, not arrays of bitvectors. Support of arrays is possible but remains future work.

\begin{proposition}
Domain propagation in bitme implements the tracker algebra of combinational-logic BTOR2 operators over bitvectors in rotor-generated models.
\end{proposition}

By tracker algebra we mean the generalization of the algebra of combinational-logic BTOR2 operators from bitvectors to trackers. Note that a tracker that maps all input values to a single state value tracks an expression that evaluates to a constant for all input values. A tracker with a Boolean codomain is another interesting special case where the tracker provides all input values that satisfy (are mapped to true) and do not satisfy (are mapped to false) the tracked Boolean expression, effectively reducing satisfiability checks to lookups.

Bitme is sound and complete on rotor-generated BTOR2 models assuming that its implementation is correct, in particular its implementation of BDDs, and in turn assuming that the SMT solvers bitme uses are also sound and complete. Assuming rotor-generated models are sound and complete on RISC-V machines makes bitme sound and complete on RISC-V machines.

\begin{proposition}
(Soundness) Given a rotor-generated RISC-V machine model $R$ and some $k\geq 0$, if bitme returns that $R$ is $k$-satisfiable ($k$-unsatisfiable), then $R$ is $k$-satisfiable ($k$-unsatisfiable).
\end{proposition}

\begin{proposition}
(Completeness) Given a rotor-generated RISC-V machine model $R$ and some $k\geq 0$, if $R$ is $k$-satisfiable ($k$-unsatisfiable), then bitme returns that $R$ is $k$-satisfiable ($k$-unsatisfiable) modulo timeout.
\end{proposition}

Just like any other bounded model checker, bitme is only as good as the bound $k$ it can explore in reasonable time. In particular, nothing can be said about machine behavior beyond $k$.

\subsection{Implementation}

Bitme is implemented in around 8k lines of Python code with minimal use of any libraries but heavy use of builtin data structures such as dictionaries (hash tables) and lists. Z3 and bitwuzla are integrated into bitme through their Python APIs. As with rotor, the code is open source under a two-clause BSD license.

Even though bitme implements algorithms whose asymptotic complexity is exponential in the size of model input, we decided to use Python in order to speed up development. We even began porting rotor into bitme but stalled that effort for now. A high-performance implementation of bitme in a performance-oriented language may nevertheless make sense eventually.

\paragraph{Model Unrolling.}

Bitme parses rotor-generated BTOR2 models into an internal representation for which bitme can generate Z3 and bitwuzla expressions that are semantically equivalent to BTOR2. Bitme also implements a BTOR2 emulator on the internal representation, similar to rotor, but extended to symbolic model input using BDDs as discussed below. Bitme can choose from a number of different mechanisms for unrolling models. When working with Z3 and bitwuzla, bitme may choose between using lambda expressions (default) and term substitution for applying transition functions. By default transition functions are generated as lambda expressions that are applied to variables representing model state in each transition. Alternatively, those variables are substituted into transition functions using term substitution provided by Z3 and bitwuzla. Another option is that bitme unroll models by duplicating transition functions internally in Python avoiding the need for lambda expressions and term substitution.

\paragraph{Algebraic Decision Diagrams.}

Bitme implements a variant of algebraic decision diagrams (ADDs) as trackers that we generalized to \emph{reduced ordered algebraic bitvector decision diagrams} (ROABVDDs) which map input values at the granularity of 8-bit bitvectors (bytes), not individual bits, to state values represented by bitvectors of any size. Implementing the ROABVDD algebra of combinational-logic BTOR2 operators essentially requires implementing the intersection and union of ROABVDDs to support binary and ternary (\texttt{ite}) BTOR2 operators, respectively. The generalization from input bits to bytes matches the input format of rotor-generated models but requires efficient handling of intersections and unions of subsets of byte-sized values. For this purpose, bitme represents subsets with 256-bit unsigned integers where intersection and union is reduced to bitwise conjunction and disjunction, respectively. The key advantage of ROABVDDs and BDDs in general is that their size is independent of any (don't-care) inputs that do not influence their output. However, while applying binary and \texttt{ite} operators requires square and linear time, respectively, in the number of involved state values, the size of BDDs such as ROABVDDs and hence the overall runtime can still be exponential in the number of inputs. As with other forms of BDDs, the order of inputs can make an exponential difference in size. Bitme uses a fixed order, namely the order in which input is read by a model. Controlling the order is interesting but remains future work. Lastly, bitme can generate a combinational-logic BTOR2 expression for a given ROABVDD that is equivalent to the ROABVDD to defer reasoning to an SMT solver whenever further domain propagation is not possible. This may happen if array access or model input that is larger than a given bound on bitvector size is encountered. We show in experiments that even partial domain propagation can speed up reasoning significantly.

\paragraph{Context-Free-Language Ordered Binary Decision Diagrams.}

Bitme also implements a variant of context-free-language ordered decision diagrams (CFLOBDDs)~\cite{sistla2024cflobdds} as trackers that we generalized to \emph{context-free-language ordered bitvector decision diagrams} (CFLOBVDDs) which map input values at the granularity of 1-bit, 2-bit, 4-bit, or 8-bit bitvectors to state values represented by bitvectors of any size. Experiments show that performance roughly doubles with each step up in granularity. The original CFLOBDD algorithm contains some code with square complexity that was easy to reduce to linear complexity, increasing performance of CFLOBVDDs further by one order of magnitude. CFLOBDDs and hence CFLOBVDDs may be exponentially smaller than reduced ordered BDDs~\cite{bryant1986graph} including ROABVDDs motivating us to integrate CFLOBDDs as CFLOBVDDs into bitme. Intuitively, CFLOBDDs may achieve the advantage by facilitating reuse of \emph{all} common subtrees in their representation of decision trees. Again, bitme can generate a combinational-logic BTOR2 expression for a given CFLOBVDD that is equivalent to the CFLOBVDD to defer reasoning to an SMT solver whenever further domain propagation is not possible.

\paragraph{Arrays.}

Arrays of bitvectors pose a performance challenge to SMT solvers. Bitme therefore supports mapping arrays of bitvectors, in particular their access logic, to individual bitvectors. Mapping arrays to individual bitvectors requires generating separate indexing logic for each individual bitvector to identify the bitvector by its index in the array, again separately for each read and each write access. Bitme can choose between iterative indexing logic that compares a given index with all indexes of the array iteratively in a linear search, and exponentially more compact recursive indexing logic that compares a given index with all indexes recursively through binary search over the bits representing indexes. Mapping arrays to bitvectors makes constant and domain propagation more effective with bitme than with rotor. Applying constant and domain propagation to arrays as is, and not just bitvectors, is possible but remains future work.

\paragraph{Space.}

Without domain propagation, bitme's spatial requirements are bounded in the number of model unrollings multiplied by the sum of the size of all combinational-logic functions, not considering the internal state of SMT solvers. With domain propagation, the bound depends on the number and size of BDDs but is independent of the number of model unrollings (as long as bitme explores control- and data-flow simultaneously breadth-first). The number of BDDs is bounded by the number of bitvectors modeling machine state (including the bitvectors of mapped arrays) plus the size of all combinational-logic functions. The size of a BDD maybe be exponential in the size of the modeled input buffer hence potentially dominating all other bounds eventually.

\paragraph{Correctness.}

Similar to rotor, bitme is complex and correctness is a major concern. Bitme can generate BTOR2 files that it has parsed to validate parser correctness. Bitme has also been validated for correctness by running it on rotor-generated models with known properties. Bitme also features a simple unit test suite that checks correctness of its implementation of CFLOBVDDs. Further validation was done by checking equivalence of ROABVDDs and CFLOBVDDs running simultaneously. However, establishing correctness of bitme formally remains a challenge.

\subsection{Usage}

\begin{table}
\centering
\begin{tabular}{|l|l|}
\hline
\textbf{console argument} & \textbf{description}\\
\hline
\texttt{-kmin k} & unroll model for k transitions skipping bad property checks up to k\\
\texttt{-kmax k} & unroll model for k transitions\\
\texttt{-{}-use-Z3} & use Z3 as solver backend\\
\texttt{-{}-use-bitwuzla} & use bitwuzla as solver backend\\
\texttt{-substitute} & use term substitution rather than lambdas in Z3 and bitwuzla\\
\texttt{-{}-use-CFLOBVDD b} & use CFLOBVDDs over $b$-bit inputs as solver backend (default: 8)\\
\texttt{-propagate p} & propagate machine input in bitvectors of size up to $p$ bits\\
\texttt{-array a} & convert bitvector arrays with index size of up to $a$ bits to bitvectors\\
\texttt{-{}-recursive-array} & use index binary search in array conversion\\
\texttt{-{}-print-pc} & print program counter in each transition\\
\texttt{-{}-check-termination} & check termination in each transition\\
\texttt{-{}-unconstraining-bad} & do not assert negations of bad properties\\
\texttt{-{}-print-transition} & print transition in each transition\\
\texttt{-{}-branching} & explore control flow depth-first\\
\hline
\end{tabular}
\caption{\texttt{bitme model\_file [output\_file] \{arguments\}}}
\label{tab:bitme-options}
\end{table}

Table~\ref{tab:bitme-options} provides a summary of bitme console arguments of which most have been described above. Bitme may also be configured to print the value of the program counter in each transition, and report each transition in order to show progress. Bitme runs even if Z3 and bitwuzla are unavailable but then relies entirely on its own implementation of BDDs.

\section{Experiments}

We now demonstrate the capabilities of rotor and bitme using a microbenchmarking suite of 18 different sample programs\footnote{see supplementary material}. Models are generated for each sample using rotor with the command line of \texttt{-riscuonly -heapallowance 128 -stackallowance 2048}, with \texttt{-bytestoread} set to the expected number of input bytes. In each case, bitme is instructed to keep exploring models until it finds a set of input bytes which would cause the program to enter a bad state; most often this takes the form of returning a bad exit code. We then measure the runtime of bitme until it finds said bad state, averaging the collected values over three runs per sample. If a given invocation of bitme does not terminate within 15 minutes (900 seconds) it is prematurely terminated and dropped from the output data set. There was no limit applied to the maximum memory usage of each run. All data was collected on a machine with a Ryzen 9950X3D CPU and 96GB of DDR5 RAM running Linux 6.12.35 (specifically NixOS 25.11 Xantusia).

First we compare the performance of SMT solvers (in this case Z3) in an SMT scenario to bitme's constant and domain propagation.
For this we invoke bitme in one of four modes:
\begin{itemize}
  \item SMT (\texttt{-propagate 0}): in this configuration bitme uses the SMT solver without applying neither constant nor domain propagation.
  \item constant propagation (\texttt{-propagate 1}\footnote{\texttt{-propagate 1} instructs bitme to turn on constant propagation as well as domain propagation for bitvectors up to one bit in size. However, since no such bitvectors occur in models generated by rotor, this effectively just turns on constant propagation in isolation.}): in this configuration bitme applies constant propagation, however domain propagation using BVDDs is not utilized.
  \item ROABVDD domain propagation (\texttt{-propagate 8 --use-ROABVDD}): in this configuration bitme applies both constant propagation as well as domain propagation using ROABVDDs for all bitvectors up to 8 bits. Since 8 bits is the size of each input bitvector (since the input is treated as a stream of bytes), higher values of \texttt{-propagate} do not have any additional effect.
  \item CFLOBVDD domain propagation (\texttt{-propagate 8 --use-CFLOBVDD}): in this configuration bitme applies both constant propagation as well as domain propagation using CFLOBDDs for all bitvectors up to 8 bits, similar to the above configuration.
\end{itemize}

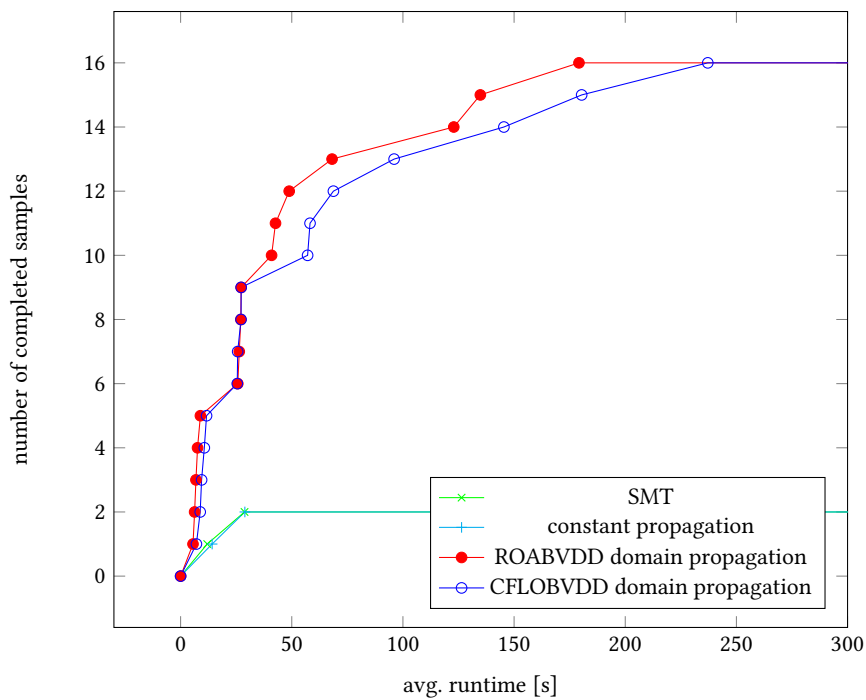
\begin{figure}[h]
  \centering
  \begin{tikzpicture}
    \begin{axis}[xmax=300,xlabel={avg. runtime [s]},ylabel={number of completed samples},legend pos=south east]
      \addplot[green,mark=x] coordinates { (0,0)  (11.977468,1) (28.750856666666664,2) (1000,2) };
      \addlegendentry{SMT};
      \addplot[cyan,mark=+] coordinates { (0,0)  (14.182913,1) (28.894724,2) (1000,2) };
      \addlegendentry{constant propagation};
      \addplot[red,mark=*] coordinates { (0,0)  (5.569718333333333,1) (6.296855,2) (6.892007333333333,3) (7.616049,4) (8.899203,5) (25.711756666666663,6) (26.379113666666665,7) (27.087002333333334,8) (27.232824333333337,9) (40.920483999999995,10) (42.68454499999999,11) (48.83933233333334,12) (68.12576233333333,13) (122.86931133333333,14) (134.81796699999998,15) (179.187892,16) (1000,16) };
      \addlegendentry{ROABVDD domain propagation};
      \addplot[blue,mark=o] coordinates { (0,0)  (7.100455333333334,1) (8.807429,2) (9.434983666666668,3) (10.660293666666666,4) (11.661524,5) (25.473632666666663,6) (25.663527000000002,7) (27.174953333333335,8) (27.230138999999998,9) (57.120737,10) (58.208945,11) (68.72782533333333,12) (96.03411433333333,13) (145.40920733333334,14) (180.39272866666667,15) (237.12627999999998,16) (1000,16) };
      \addlegendentry{CFLOBVDD domain propagation};
    \end{axis}
  \end{tikzpicture}
  \caption{Bitme solver performance: nearly all samples require domain propagation to finish executing within the maximum allocated runtime of 15 minutes (runtime in seconds, lower is better)}
  \label{fig:sat-vs-prop}
\end{figure}

Fig.~\ref{fig:sat-vs-prop} shows the collected data. Note how a large number of sample programs are only able to finish executing when domain propagation is being used. This is a result of the 15 minute timeout mentioned at the start of the section; said samples are only able to succesfully finish within the alloted time frame when domain propagation is active. Additionally, one of the two samples which are able to finish executing in an SMT setting (\texttt{division-by-zero-3-35.c}) similarly sees its runtime reduced when domain propagation is turned on; the other sample however (\texttt{memory-access-fail-1-35.c}) does not finish at all, which represents an instance of the application of BVDDs actually introducing a performance regression. As such this demonstrates the performance advantages domain propagation using BVDDs can obtain compared to standard SMT solvers. CFLOBVDDs however seem to not be able to obtain as strong of an result as ROABVDDs, with samples seemingly taking a nearly-constant amount of time longer to execute. This can mostly be attributed to the increased implementation overhead of CFLOBVDDs; CFLOBVDDs have been proven to not perform "significantly worse" compared to ROABVDDs as an upper bound\cite{zhi2025polynomial}.

We now briefly turn our attention to the \texttt{-array} parameter. This parameter instructs bitme to convert arrays up to a certain size (specified as the number of index bits / the base-2 logarithm of the array size) to a set of individual flattend bitvectors.

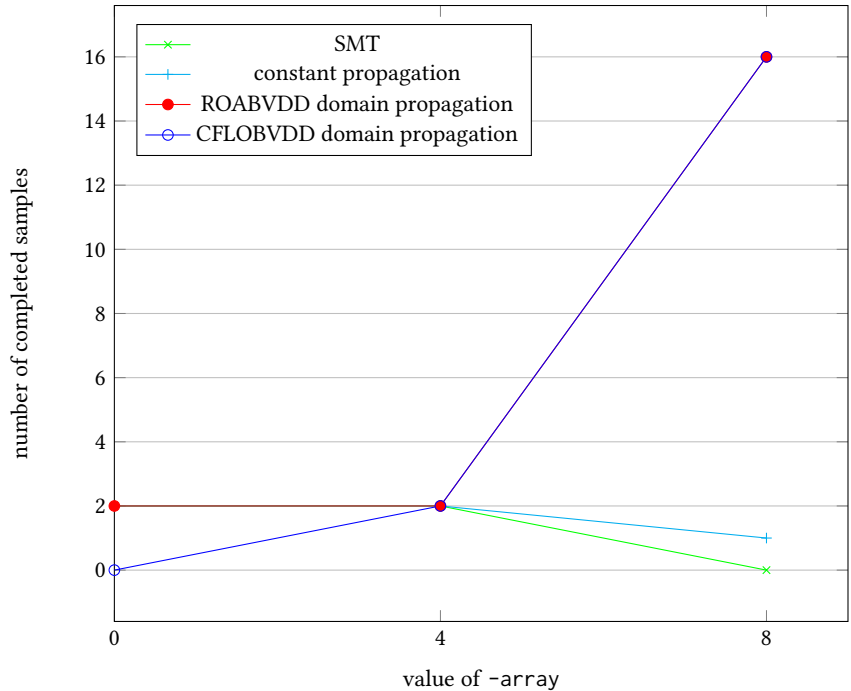
\begin{figure}[h]
  \centering
  \begin{tikzpicture}
    \begin{axis}[xlabel={value of \texttt{-array}},ylabel={number of completed samples},ymajorgrids,xtick={0,4,8},xmin=0,xmax=9,legend pos=north west]
      \addplot[green,mark=x] coordinates { (0,2) (4,2) (8,0) };
      \addlegendentry{SMT};
      \addplot[cyan,mark=+] coordinates { (0,2) (4,2) (8,1) };
      \addlegendentry{constant propagation};
      \addplot[red,mark=*] coordinates { (0,2) (4,2) (8,16) };
      \addlegendentry{ROABVDD domain propagation};
      \addplot[blue,mark=o] coordinates { (0,0) (4,2) (8,16) };
      \addlegendentry{CFLOBVDD domain propagation};
    \end{axis}
  \end{tikzpicture}
  \caption{Number of completed samples across various values of \texttt{-array}: only two samples finish executing without domain propagation. CFLOBVDDs do not finish executing any samples with \texttt{-array 0} due to limitations in implementation.}
  \label{fig:array-cnt}
\end{figure}

It can be seen in the above figures that in the non-BVDD/domain-propagation modes, the performance of bitme actually decreases as more and more arrays are unfolded, eventually resulting in less samples being able to finish executing within the timeout. We assume this is because SMT solvers built-in handling of arrays is more tightly integrated with the rest of the solver, allowing it to reason more clearly about input problems. However, for domain-propagation to be effective, a value of \texttt{-array 8} is required. This is due to an internal limitation of bitme which prevents it from applying domain-propagation to values involving non-unfolded arrays. This limitation may be addressed in future works. For past tests we have utilized \texttt{-array 0} for the non-domain propagation modes, and \texttt{-array 8} for invocations utilizing domain propagation. CFLOBVDDs inability to produce any samples for \texttt{-array 0} is due to the result of a similar internal limitation.

Various rotor arguments also produce slightly different models, which subsequently change bitme's performance characteristics. As such we will now be comparing the performance of models generated using no special options (which generates models targetting the full RISC-V architecture), using \texttt{-riscuonly} (the default so far, which generates models targeting the subset of RISC-V generated by selfie's compiler).

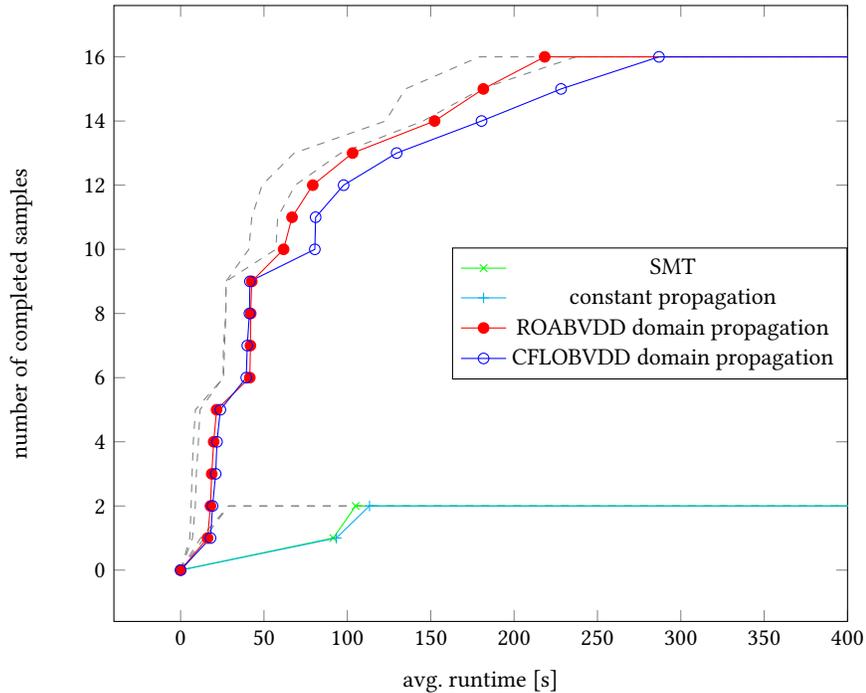
\begin{figure}[h]
  \centering
  \begin{tikzpicture}
    \begin{axis}[xmax=400,xlabel={avg. runtime [s]},ylabel={number of completed samples},legend style={at={(1,0.5)}, anchor=east}]
      \addplot[dashed,no markers,color=gray,forget plot] coordinates { (0,0)  (11.977468,1) (28.750856666666664,2) (1000,2) };
      \addplot[dashed,no markers,color=gray,forget plot] coordinates { (0,0)  (14.182913,1) (28.894724,2) (1000,2) };
      \addplot[dashed,no markers,color=gray,forget plot] coordinates { (0,0)  (5.569718333333333,1) (6.296855,2) (6.892007333333333,3) (7.616049,4) (8.899203,5) (25.711756666666663,6) (26.379113666666665,7) (27.087002333333334,8) (27.232824333333337,9) (40.920483999999995,10) (42.68454499999999,11) (48.83933233333334,12) (68.12576233333333,13) (122.86931133333333,14) (134.81796699999998,15) (179.187892,16) (1000,16) };
      \addplot[dashed,no markers,color=gray,forget plot] coordinates { (0,0)  (7.100455333333334,1) (8.807429,2) (9.434983666666668,3) (10.660293666666666,4) (11.661524,5) (25.473632666666663,6) (25.663527000000002,7) (27.174953333333335,8) (27.230138999999998,9) (57.120737,10) (58.208945,11) (68.72782533333333,12) (96.03411433333333,13) (145.40920733333334,14) (180.39272866666667,15) (237.12627999999998,16) (1000,16) };
      \addplot[green,mark=x] coordinates { (0,0)  (91.55317666666667,1) (104.97158566666667,2) (1000,2) };
      \addlegendentry{SMT};
      \addplot[cyan,mark=+] coordinates { (0,0)  (93.35625733333332,1) (113.29823133333332,2) (1000,2) };
      \addlegendentry{constant propagation};
      \addplot[red,mark=*] coordinates { (0,0)  (16.150443,1) (17.900296,2) (18.658265,3) (19.827598333333334,4) (21.579945999999996,5) (41.502643,6) (41.87057433333333,7) (42.052002333333334,8) (42.676244,9) (61.778399,10) (66.83002966666668,11) (79.24851,12) (103.12608633333333,13) (152.31603033333332,14) (181.54859299999998,15) (218.35902066666665,16) (1000,16) };
      \addlegendentry{ROABVDD domain propagation};
      \addplot[blue,mark=o] coordinates { (0,0)  (17.936584,1) (19.153799000000003,2) (21.033970333333333,3) (21.80108733333333,4) (23.84578833333333,5) (39.36912233333333,6) (39.962160000000004,7) (41.27282266666666,8) (41.571618666666666,9) (80.49442033333334,10) (80.97381233333333,11) (97.76852566666666,12) (129.564624,13) (180.47943766666666,14) (228.172944,15) (286.88438233333335,16) (1000,16) };
      \addlegendentry{CFLOBVDD domain propagation};
    \end{axis}
  \end{tikzpicture}
  \caption{Bitme solver performance: on rotor models generated for the full RISC-V instruction set. Previously collected performance data is also displayed as a baseline (runtime in seconds, lower is better)}
  \label{fig:full-riscv}
\end{figure}

As can be seen in Fig.~\ref{fig:full-riscv}, SMT solvers are affected in much greater capacity from the slightly increased model complexity resulting from targeting the full RISC-V architecture compared to BVDDs, which only take a nearly constant hit in performance.

We have so far demonstrated that domain propagation is able to improve upon the runtime of regular SMT solvers. We will now compare domain propagation using CFLOBVDDs to ROABVDDs. As mentioned in earlier work CFLOBVDDs have the potential to further improve upon ROABVDDs in terms of scalability because of their ability to reuse parts of the decision tree across different branches. To effectively demonstrate this, we focus our attention on two new samples designed to leverage this ability, named \texttt{cflobvdd-multi-input-X.c} and \texttt{cflobdd-bit-inversion-X.c}. We measure both the average runtime as well as the maximum memory usage throughout each run.

\texttt{cflobvdd-multi-input-X.c} is a simple program which consumes X input bytes, keeping count of how many of them match the character \texttt{0}. All X input bytes must match this character for the state to be considered ``bad''. This is designed to test the ability of CFLOBVDDs to reuse parts of the decision tree when confronted with an increasing number of input variables. \texttt{cflobdd-bit-inversion-X.c} on the other hand reads a single input digit, then flips the order of the first X bits of the value read. This is designed to test CFLOBVDDs in regards to symbolic branches---branches whose condition depends on some symbolic state. The more bits of the input value are flipped, the more complex the tree structure of the corresponding result BVDD has to get.

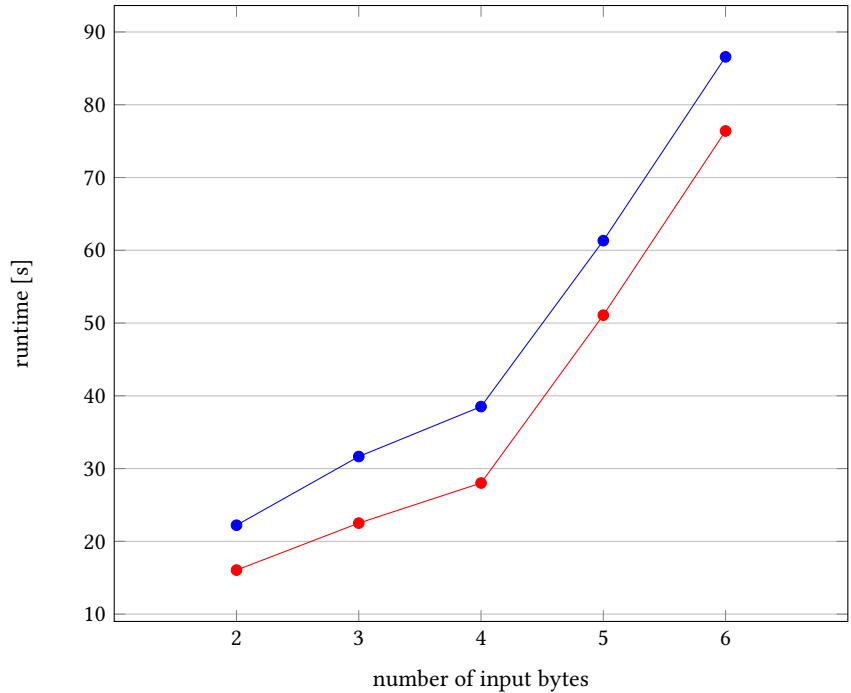
\begin{figure}[h]
  \centering
  \begin{tikzpicture}
    \begin{axis}[xlabel={number of input bytes},ylabel={runtime [s]},ymajorgrids,xtick={2,3,4,5,6},xmin=1,xmax=7]
    \addplot[color=red,mark=*] coordinates {(2,16.048532333333334) (3,22.507159333333334) (4,28.015620333333334) (5,51.08170466666667) (6,76.39505166666667) };
    \addplot[color=blue,mark=*] coordinates {(2,22.21864633333333) (3,31.656125666666668) (4,38.51964866666666) (5,61.321812333333334) (6,86.58188533333333) };
    \end{axis}
  \end{tikzpicture}
  \caption{Average runtime of \texttt{cflobvdd-multi-input-X.c} using ROABVDD/CFLOBVDD-based domain propagation (marked in red/blue respectively) over increasing values of X (runtime in seconds, lower is better)}
  \label{fig:cflobvdd-multiinput-run}
\end{figure}
\begin{figure}[h]
  \centering
  \begin{tikzpicture}
    \begin{axis}[xlabel={number of input bytes},ylabel={memory usage [MB]},ymajorgrids,xtick={2,3,4,5,6},xmin=1,xmax=7]
    \addplot[color=red,mark=*] coordinates {(2,1055.90234375) (3,1343.6861979166667) (4,1551.42578125) (5,2619.4895833333335) (6,3749.4596354166665) };
    \addplot[color=blue,mark=*] coordinates {(2,612.7903645833334) (3,692.421875) (4,753.26171875) (5,897.4700520833334) (6,1044.4622395833333) };
    \end{axis}
  \end{tikzpicture}
  \caption{Average maximum memory usage of \texttt{cflobvdd-multi-input-X.c} using ROABVDD/CFLOBVDD-based domain propagation (marked in red/blue respectively) over increasing values of X (memory usage in MBs, lower is better)}
  \label{fig:cflobvdd-multiinput-mem}
\end{figure}
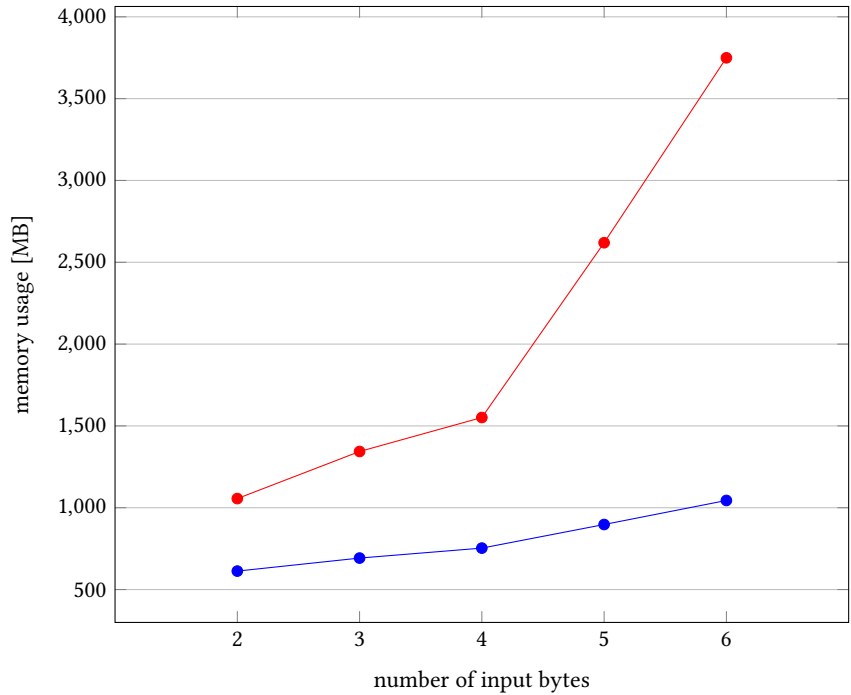

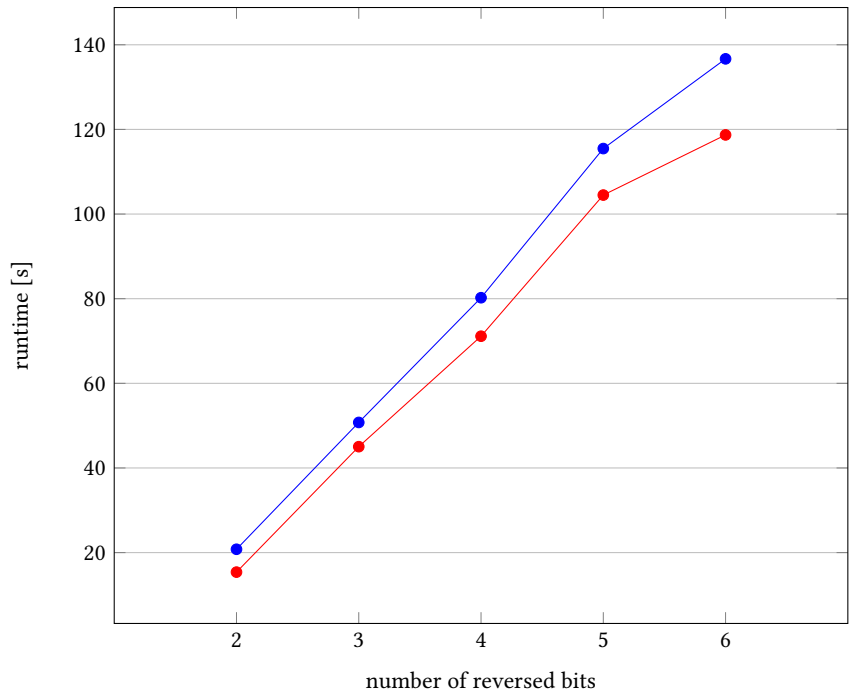
\begin{figure}[h]
  \centering
  \begin{tikzpicture}
    \begin{axis}[xlabel={number of reversed bits},ylabel={runtime [s]},ymajorgrids,xtick={2,3,4,5,6},xmin=1,xmax=7]
    \addplot[color=red,mark=*] coordinates {(2,15.388053666666666) (3,45.03178466666666) (4,71.13416833333333) (5,104.48757966666666) (6,118.70216066666667) };
    \addplot[color=blue,mark=*] coordinates {(2,20.788877666666664) (3,50.757531666666665) (4,80.22934233333332) (5,115.47487466666666) (6,136.675578) };
    \end{axis}
  \end{tikzpicture}
  \caption{Average runtime of \texttt{cflobvdd-bit-inversion-X.c} using ROABVDD/CFLOBVDD-based domain propagation (marked in red/blue respectively) over increasing values of X (runtime in seconds, lower is better)}
  \label{fig:cflobvdd-bitinv-run}
\end{figure}
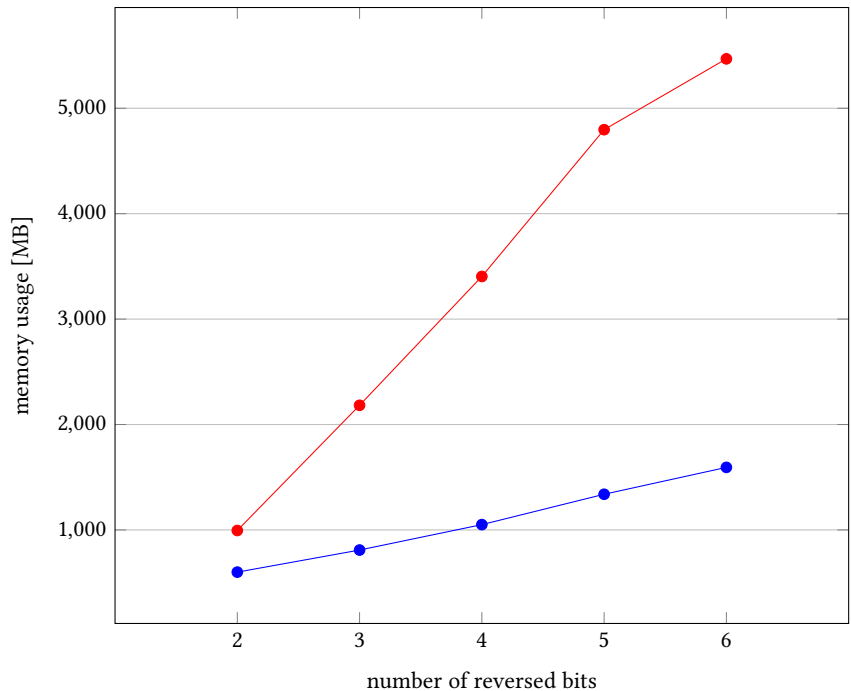
\begin{figure}[h]
  \centering
  \begin{tikzpicture}
    \begin{axis}[xlabel={number of reversed bits},ylabel={memory usage [MB]},ymajorgrids,xtick={2,3,4,5,6},xmin=1,xmax=7]
    \addplot[color=red,mark=*] coordinates {(2,994.6484375) (3,2182.3802083333335) (4,3404.3697916666665) (5,4796.7421875) (6,5468.690104166667) };
    \addplot[color=blue,mark=*] coordinates {(2,599.7317708333334) (3,809.2760416666666) (4,1050.3450520833333) (5,1338.35546875) (6,1593.4244791666667) };
    \end{axis}
  \end{tikzpicture}
  \caption{Average maximum memory usage of \texttt{cflobvdd-bit-inversion-X.c} using ROABVDD/CFLOBVDD-based domain propagation (marked in red/blue respectively) over increasing values of X (memory usage in MBs, lower is better)}
  \label{fig:cflobvdd-bitinv-mem}
\end{figure}

It can be seen in Figs.~\ref{fig:cflobvdd-multiinput-run} and \ref{fig:cflobvdd-bitinv-run} that while ROABVDDs still outperform CFLOBVDDs in terms of their runtime, CFLOBVDDs tend to perform significantly better in the metric of memory usage, which also tends to scale linearly as compared to ROABVDDs whose maximum memory usage scales in-step with its runtime (see Figs.~\ref{fig:cflobvdd-multiinput-mem} and \ref{fig:cflobvdd-bitinv-mem}). This marks CFLOBVDDs as a promising tool to be used for bounded-model checking increasingly complex programs compared to standard ROABVDDs. We expect that the small nearly constant gap in performance between the two types of BVDD can be closed further by utilizing a more efficient implementation of CFLOBVDDs.

\section{Future Work}

Designing and implementing rotor and bitme has been a major multi-year effort and yet making a breakthrough in scalable bit-precise reasoning over software remains an open problem. There are numerous directions for future work based on rotor and bitme that may be worthwhile pursueing. Promising engineering challenges are:

\begin{enumerate}
\item integrating rotor into bitme to reason about RISC-V machine code with a single tool,
\item implementing bitme in performance-oriented languages such as Rust or C++,
\item support of floating-point arithmetic and other RISC-V extensions,
\item support of more system calls and additional failure modes,
\item support of other forms of model input such as symbolic console arguments,
\item trading off pre-fetching and pre-decoding between model generation and model checking time, and
\item establishing rotor as benchmarking platform for SMT solvers.
\end{enumerate}

More principled venues for interesting future work include:

\begin{enumerate}
\item applying BDDs to arrays of bitvectors rather than mapping arrays to individual bitvectors,
\item support of checks for block-level memory safety and liveness properties,
\item developing CFLOBDDs further into a bit-precise reasoning backend for compositional symbolic execution that scales by identifying functional behavior of machine code,
\item formal verification of rotor and bitme correctness, and
\item support of concurrent code, program equivalence checking, and code synthesis in bitme.
\end{enumerate}

\section{Conclusions}

We have developed two tools: rotor, for generating bit-precise models of RISC-V machines, and bitme, for reasoning over rotor-generated models using bounded model checking and domain propagation. Bitme utilizes Z3 and Bitwuzla as well as reduced ordered algebraic bitvector decision diagrams (ROABVDDs) and context-free-language ordered bitvector decision diagrams (CFLOBVDDs). Experiments show that domain propagation, particularly with ROABVDDs, significantly boosts performance in bounded model checking of RISC-V machine code, outperforming traditional SMT solvers. Although ROABVDDs currently show faster absolute runtimes, CFLOBVDDs exhibit superior scalability with increasing input complexity in regards to their memory usage, with their associated runtime overhead being effectively constant. Experiments also show that while unfolding arrays can hinder performance in non-BVDD/domain-propagation modes due to less integrated array handling in SMT solvers, it is essential for effective domain propagation. Future work will focus on integrating tools, improving performance with different programming languages, expanding RISC-V feature support, optimizing model generation and checking, formally verifying correctness, and enhancing BDD application to arrays.

\section{Acknowledgments}

This work was co-funded by the Czech Science Foundation under Grant No. 23-07580X and the European Union under the project Robotics and Advanced Industrial Production (reg.~no.~CZ.02.01.01/00/22\_008/0004590).

\bibliographystyle{ACM-Reference-Format}
\bibliography{paper}

\end{document}